\documentclass[11pt,twoside]{article}
\usepackage{./asp2014}

\newcommand{\ls}{\mbox{$L_{\odot}$}}
\newcommand{\ms}{\mbox{$M_{\odot}$}}

\newcommand{\gsim}{\raisebox{-.4ex}{$\stackrel{>}{\scriptstyle \sim}$}}
\newcommand{\lsim}{\raisebox{-.4ex}{$\stackrel{<}{\scriptstyle \sim}$}}

\newcommand{\pasa}{Pub. Astron. Soc. of Australia}
\newcommand{\rmxaa}{RMxAA}
\newcommand{\captionfonts}{\small}

\makeatletter  
\long\def\@makecaption#1#2{%
  \vskip\abovecaptionskip
  \sbox\@tempboxa{{\captionfonts #1: #2}}%
  \ifdim \wd\@tempboxa >\hsize
    {\captionfonts #1: #2\par}
  \else
    \hbox to\hsize{\hfil\box\@tempboxa\hfil}%
  \fi
  \vskip\belowcaptionskip}
\makeatother   
                                                    
\aspSuppressVolSlug
\resetcounters

\begin{document}

\title{Science with an ngVLA: Probing Strong Binary Interactions and Accretion in AGB stars with the ngVLA}
\author{Raghvendra Sahai$^1$
\affil{$^1$JPL, Pasadena, CA, USA; \email{raghvendra.sahai@jpl.nasa.gov}}}

\paperauthor{Raghvendra Sahai}{raghvendra.sahai@jpl.nasa.gov}{ORCID_Or_Blank}{JPL}{Astrophysics \& Space Sciences}{Pasadena}{CA}{91109}{USA}

\abstract
Understanding strong binary interactions is of wide astrophysical importance, and the deaths of 
most stars in the Universe that evolve in a Hubble time could be fundamentally affected by such interactions. 
These stars end their lives, evolving from Asymptotic Giant Branch stars with extensive mass-loss 
into planetary nebulae with a spectacular array of morphologies. Binarity, 
and the associated formation of accretion disks (that drive collimated, fast jets) during the very late AGB or early post-AGB phase 
is believed to produce this 
dramatic morphological transformation. But the evidence for binarity and accretion during the AGB phase 
has been hard to obtain due to observational limitations. However, recent observations at UV and X-ray 
wavelengths have broken thru the observational barrier -- 
our studies using GALEX reveal a candidate population of AGB stars, generally with 
strongly-variable far-ultraviolet (FUV) emission (fuvAGB stars), and our follow-up studies with XMM-Newton, 
Chandra, and HST of a few key objects supports our hypothesis that these objects have companions 
that are actively accreting material from the primary. The most prominent fuvAGB 
star has been detected with the VLA, showing the presence of variable non-thermal emission. 
The ngVLA, with its unprecedented 
sensitivity, is needed to survey a statistical sample of fuvAGB stars over the $\sim$3-90\,Ghz range to search 
for and characterize the nature of the radio emission from fuvAGB stars and test our binarity+accretion 
hypothesis. Such a survey will distinguish 
between binarity/accretion-related radio emission that is 
expected to have both thermal and non-thermal components and display significantly time-variable on short time-scales 
(minutes to weeks), and single-star chromospheric emission from the primary that is expected to be thermal, possibly with time-variability, but only on 
long time-scales (many months to a year).

\section{Introduction -- Binarity and the AGB to PN Transition}
A major astrophysical accomplishment of the 20th century was the success of models of the 
structure and evolution of stars of all masses from adolescence to retirement. The challenges continuing 
into the 21st century are now stellar birth, stellar 
death, and the impact of binary interactions on stellar evolution. Binary interactions in particular affect our  
understanding of stars in general, since such interactions affect 
the observable qualities of stars and stellar populations.

A glaring example of the effect of binary interactions on stellar evolution are planetary nebulae (PNe) 
that represent the bright end-stage of most stars in the Universe that evolve in a Hubble time (i.e., those with 
main-sequence masses of $1-8$\,\ms). PNe have wide-ranging astrophysical importance, covering diverse topics such as 
mass-loss and its effect on stellar evolution, the chemical evolution of the ISM, the cosmological distance ladder, 
astrophysical jets, common-envelope evolution, and intermediate-luminosity transients. 

A long-standing puzzle for PNe formation is that while modern imaging surveys indicate that the vast majority of PNe 
deviate strongly from spherical symmetry (e.g., Schwarz, Corradi \& Melnick 1992, Manchado et al. 1996, Sahai \& Trauger 
1998 [ST98], Sahai et al. 2011a, Stanghellini et al.\,2016), the progenitors of PNe, AGB stars, have 
spherically-symmetric circumstellar envelopes (CSEs) resulting from mass-loss. Spherical mass-loss on the AGB implictly 
implies that potential physical causes of asymmetry present during the main-sequence phase (e.g., stellar rotation, 
which is very small and can be ignored at ages $> few \times 10^8$ yr) are not significant in affecting mass-loss on the 
AGB and beyond. Binarity provides a source of angular momentum, as well as a preferred axis to a stellar system, and is 
now widely believed to dramatically affect the evolutionary transition from the AGB to the planetary nebula (PN) phase 
(e.g., Balick \& Frank 2002). ST98 proposed that highly-collimated, fast jet-like outflows at the late-AGB 
or pre-PN (PPN) phase sculpt the AGB mass-loss envelopes from the inside-out, producing the observed 
aspherical morphologies of PPNe (Sahai et al. 2007) and PNe. The engines that drive these outflows must reside 
in accretion disks that can be produced as a result of binarity and resulting accretion modes due to, e.g., Bondi-Hoyle or Roche-lobe 
accretion or during common-envelope evolution (e.g., Blackman \& Lucchini 2014).

If binarity plays a significant role in the PN population, one may question whether the PN population is 
representative of all $1-8$\,\ms~stars, and one may need to significantly revise how PN are used as problems of stellar 
and chemical evolution (e.g., De Marco \& Izzard\,2017, Kwitter et al. 2014, Akashi \& Soker\,2013, 
Ciardullo et al.\,2002). Hence, obtaining observational constraints on 
binarity and its effects in the general population of AGB stars is vital. 

\subsection{Finding Binaries in AGB Stars -- An Observational Challenge} 
\begin{figure}
\vspace{-0.2in}
\resizebox{1.2\textwidth}{!}{\includegraphics{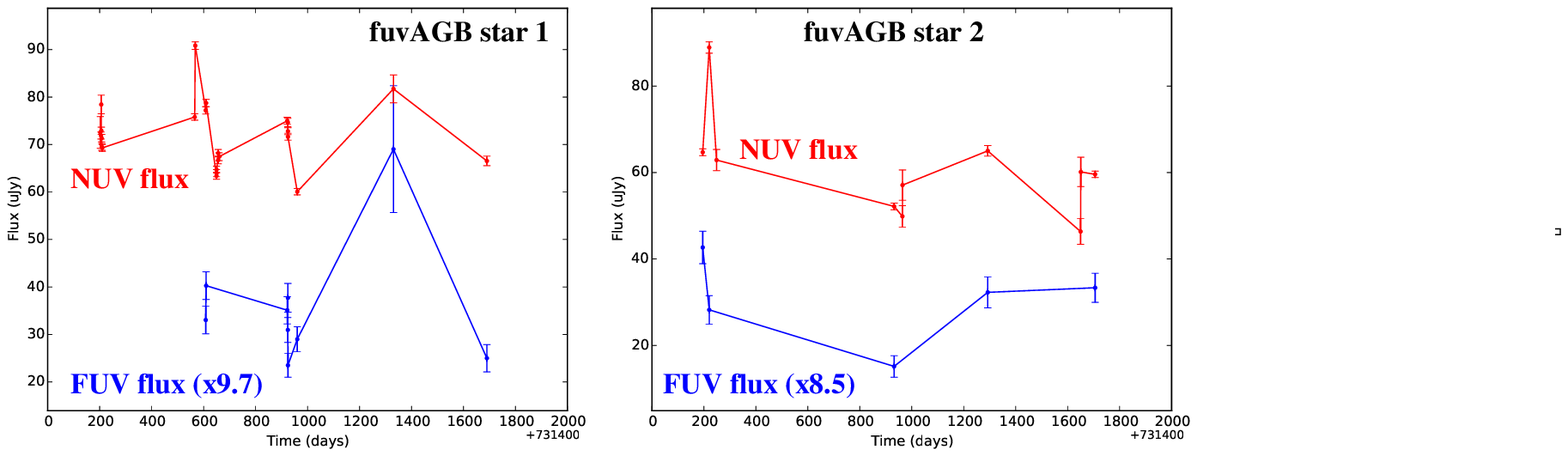}}
\vspace{-0.1in}
\caption{Two fuvAGB stars observed with GALEX showing strong variability in their FUV (blue) and NUV (red) emission (the FUV fluxes have been scaled up 
by arbitrary factors.)}
\label{fuvagbvar}
\end{figure}

Until recently, observational evidence of 
binarity in AGB stars had been sorely lacking simply because AGB stars are very luminous and variable, invalidating 
standard techniques for binary detection (e.g., radial-velocity and photometric variations due to a companion star, 
direct imaging.) An innovative technique to search for FUV and NUV emission in AGB stars with GALEX has now provided a 
large candidate sub-class of AGB stars with close binary companions (Sahai et al. 2008 [Setal08], Sahai et al.\,2011b [Setal11], Sahai et al. \,2016). In this 
sub-class, for objects with emission in the FUV (1344-1786\AA) (fuvAGB stars), the observed FUV fluxes are typically a 
factor $>10^{6}$ larger than expected for the primary's photospheric emission (Setal08), and show strong 
variability (Fig.\,\ref{fuvagbvar}). 

It should be noted that fuvAGB stars in general (based on their optical 
spectra), do not belong to the well-studied class of symbiotic stars (red giant stars with white-dwarf [WD] companions) and 
have never been classified as such. So if the compact companions in fuvAGB star systems are WDs, then these 
must be quite cool ($T_{eff}\lesssim20,000$\,K) (Sahai et al\,2015 [Setal15]). 

\subsection{UV \& X-Ray Emission: Binarity+Accretion and/or Chromospheres?} 
\begin{figure}[htb]
\hspace{-0.3in}
\resizebox{1.05\textwidth}{!}{\includegraphics{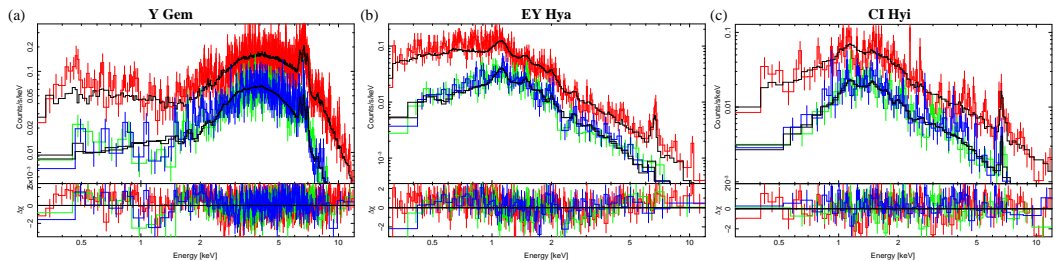}}
\vskip -0.1in
\caption{X-ray spectra fuvAGB stars Y\,Gem, EY\,Hya and CI\,Hyi using XMM/EPIC detectors pn ({\it red}),
MOS1 ({\it blue}), MOS2 ({\it green}), together with
APEC model fits ({\it black}); bottom panels show residuals.
}
\label{xrayspec}
\end{figure}

\begin{figure}[htb]
\hskip -0.05in
\resizebox{1.0\textwidth}{!}{\includegraphics{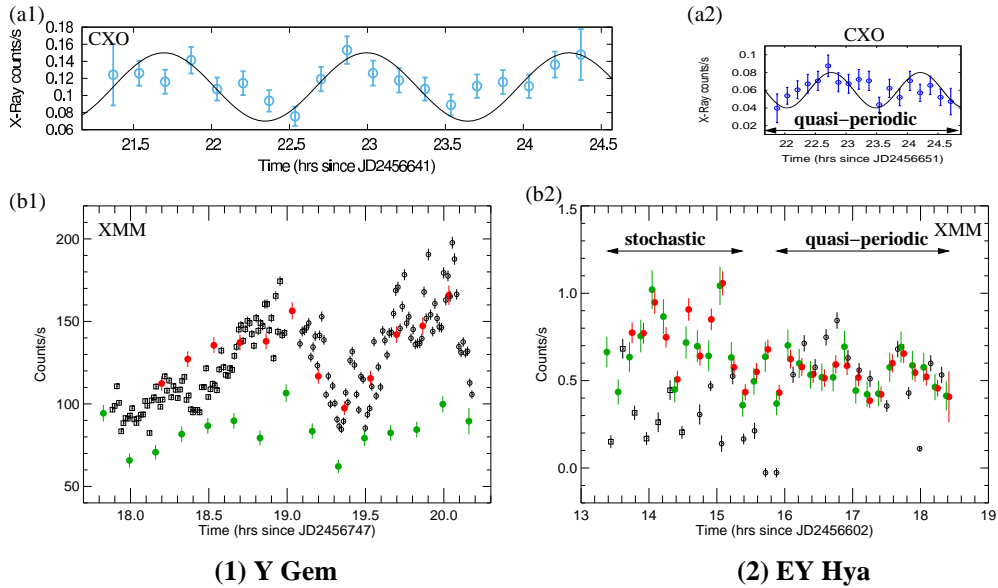}}
\caption{X-ray and UV variability in the fuvAGB stars Y\,Gem and EY\,Hya. ({\it a}) CXO (ACIS-S), and
({\it b}) XMM (EPIC=pn+MOS1+MOS2:\,red, MOS=MOS1+MOS2:\,green, UVM2:\,black squares, UVW2:\,black circles).
The EPIC, MOS and UVW2 data have been
respectively re-scaled as follows: 80,\,130,\,2.5\,(Y\,Gem), and 2,\,5,\,0.8\,(EY\,Hya). A sinusoidal fit
(by eye) with periods, P=1.35 and 1.45\,hr are shown for the X-ray light curves of Y\,Gem and EY\,Hya in panels {\it
b1}
\& {\it b2}.}
\label{lc-xray-uv}
\vskip -0.01in
\end{figure}

The FUV source in fuvAGB stars has been hypothesized to be likely dominated by emission due to variable accretion 
activity associated with a close companion (Setal08, Setal11, Ortiz \& Guerrero 2016). 
Small X-rays surveys using XMM-Newton and Chandra support this hypothesis, finding X-ray emission in about 50\% of 
fuvAGB stars. The X-ray emission is characterised by relatively high luminosities $Lx\sim(0.002-0.11)$\,\ls, and very 
high plasma temperatures $Tx\sim(35-160)\times10^6$\,K (Fig.\,\ref{xrayspec}, hereafter Setal15). 
Amongst fuvAGB stars, objects with large FUV/NUV ratios, $R_{fuv/nuv}>0.2$, have a much higher probability of being 
detected in X-ray emission (Sahai et al\,2016), and are almost certainly binaries with accretion activity powering the 
high-energy emission. The UV and X-ray emission is variable (Fig.\,\ref{lc-xray-uv}) with both periodic and stochastic 
components, signatures of active, ongoing accretion (Setal15).

A recent STIS spectroscopic study of the prototype high FUV/NUV ratio star, Y Gem, shows the presence of flickering 
on time-scales of $\lesssim$20\,s and 
high-velocity infall and outflows, and thus directly supports the binary/accretion hypothesis (Sahai et al.\,2018). 
Setal11 discuss five plausible models for the UV emission from Y\,Gem and conclude that the most plausible 
one is line and/or continuum emission associated with variable accretion activity and a disk around a companion 
star -- the accretion shock may reside on (a) the disk or (b) the stellar surface of the companion (as, e.g., in T Tauri 
stars: Calvet \& Gullbring 1998). Further support for this model comes from Y\,Gem's unusually narrow CO J=2-1 line 
profile (Setal11) that suggests the possibility of a large, orbiting reservoir of molecular material in this 
object that may be associated with the central accretion disk.

This object (and objects like it) may represent the earliest phases of an AGB star with a growing accretion 
disk which will ultimately produce collimated jets that are now believed to sculpt the round circumstellar envelopes of 
AGB stars into bipolar planetary nebulae. Following the discovery of Y\,Gem's remarkable UV emission with this technique 
(Setal11), we recently (Dec 2012) obtained multifrequency (5.5, 22, \& 30.5 GHz) observations of 
Y\,Gem, detecting the source at all 3 frequencies. These data suggest that we are probing an ionized accretion disk with 
magnetospheric accretion in this object. New VLA observations of Y\,Gem, covering a more extensive wavelength space, as 
well as multiple epochs, are needed to test this scenario.

Objects with little or no FUV emission, i.e., with $R_{fuv/nuv} \lsim 0.1$, which dominate the 
population of UV-emitting AGB stars, the UV emission may have a different source. From an analysis 
of the NUV emission in 179 AGB stars, Montez et al. (2017) argue that the origin of the 
GALEX-detected UV emission is intrinsic to the AGB star (chromospheric \& photospheric emission), 
and is unrelated to binarity. Ortiz \& Guerrero (2016), from a study of  
a volume-limited sample ($<$0.5\,kpc) of 58 AGB stars, conclude that the detection of 
NUV emission with a very large observed-to-predicted ratio, $Q_{NUV}>20$ (Ortiz \& Guerrero 2016), is evidence for binarity in these 
objects.

\section{Radio Observations}
Our current working hypothesis for UV emission from AGB stars is that objects with a close companion produce high 
FUV/NUV ratios at least some of the time (since accretion can be variable) and/or a very large NUV excess, whereas 
single AGB stars (or those with large binary separations) always show low FUV/NUV ratios, as the emission in 
these is chromospheric. High-sensivity radio observations offer an unprecedented opportunity for testing the above hypothesis 
and leading to an understanding of binarity 
and binary interaction in AGB stars by observing the radio emission from AGB stars with UV emission. We describe below 
our VLA study of Y\,Gem (Sahai \& Claussen 2018, {\it in prep}) that motivates and demonstrates this concept.

\begin{figure}[hbt]
\vspace{-0.15in}
\resizebox{1.0\textwidth}{!}{\includegraphics{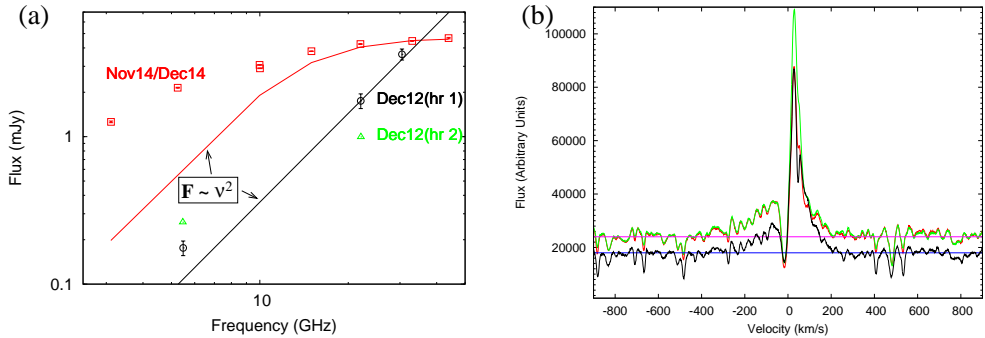}}
\vspace{-0.05in}
\caption{(a) Variable radio emission (VLA data) from Y\,Gem.
Free-free emission components have been fitted to the data at K (22\,GHz) band and higher
frequencies. The lower-frequency data shows significant
excess emission above these components at both epochs, implying the
presence of a variable component that may be due to gyrosynchrotron
radiation from material in an accretion flow. (b) Y\,Gem's H$\alpha$ line profile at 3 epochs
1\,Apr\,2012 (red), 4\,Apr\,2012 (green), and 11\,May\,2012
(black) showing a P-Cygni type profile with broad wings, signature of
a fast outflow -- the profile is time-variable.
}
\vskip -0.05in
\label{ygem-vla}
\end{figure}

Our VLA A-array observation in Dec 2012 of Y\,Gem revealed an unresolved radio source with fluxes of 
$0.176\pm0.02$\,mJy, $1.75\pm0.20$\,mJy, and $3.62\pm0.31$\,mJy at 5.5, 22 and 30.5\,GHz, respectively 
(Fig.\ref{ygem-vla}). These fluxes are far larger (factor $\sim$100) than that expected from photospheric emission, and increased 
dramatically in Nov--Dec 2014 (by a factor $\gsim$10 at $\nu<10$\,GHz). In Dec 2012, the 
spectral-index, $\alpha$, is $\sim2$ in the (22--30.5)\,Ghz, and $\sim1.65$ in the (5.5--30.5)\,GHz range, arguing 
against the emission arising in an ionized outflow, which would result in $\alpha\sim0.6$. Interpreting the 
(22--30.5)\,GHz SED as optically-thick emission from ionized gas (component 1), we find that there is a significant 
excess above such emission at 5.5\,GHz, clearly requiring an additional source (component 2) at the low-frequency end -- 
a plausible mechanism for this is gyrosynchrotron radiation. Modeling of component 1 shows that it arises in a region of 
size $\theta<0.03{''}$ (17.5\,AU at 580\,pc), with an average electron density, $n_e>8\times10^6$\,cm$^{-3}$. In Nov--Dec 2014, 
the non-thermal emission is much stronger and easily visible at low-frequencies ($\nu<10$\,GHz).

Plasma trapped in a magnetic field in the vicinity of the primary AGB star and/or the accretion 
disk may be the source of the non-thermal radio emission (component 2) from Y\,Gem -- the presence of 
such plasma is revealed by the detection of strong ($L_x\sim0.049$\,\ls) X-ray emission from Y\,Gem with Chandra (Setal15). A  
model fit to the spectrum shows the presence of plasma with a very high temperature, $T_x\sim5.4$\,keV, suggesting that 
the X-rays come from coronal emission due to plasma confined  by a strong magnetic field.

Y\,Gem's UV and X-ray variability indicates episodic accretion (a) onto the disk, from matter ejected by the AGB star (e.g.,
Mastrodemos \& Morris 1998), or (b) onto the companion from the inner disk region. If the mechanism is type $a$ then
one might expect shock and associated thermal emission, and if type $b$ and if the companion has a strong magnetic field,
then one would expect gyroresonance or gyrosynchrotron emission at radio wavelengths. Y\,Gem's radio fluxes are 
consistent with those observed in stellar sources with known accretion activity: 

For example, in the type $a$ source Mira (a well-studied AGB star which shows variable UV
line emission, and is a known (symbiotic) binary), the compact companion accretes matter from the primary's wind. The radio
fluxes of Mira (scaled to Y\,Gem's distance and factor 43 higher FUV flux, at its peak UV-emission state) of 0.38, 1.2 \&
1.9\,mJy at 8.5, 22.5, \& 43.3 GHz, are comparable to those of Y\,Gem. The UV variability of Mira results from changes 
in the accretion rate onto the companion (Matthews \& Karovska 2006). 

We note that Y\,Gem is not a symbiotic star, and also consider type $b$ sources, namely other
radio-variable stellar sources with accretion flows. Y\,Gem's radio fluxes are also consistent with emission from such 
a flow. An
example is the non-thermal (likely) gyrosynchrotron emission source observed in T\,Tau S at 2--3.6\,cm. This emission, which
may be due to a scaled-up solar-like flare or accretion-related (Skinner \& Brown 1994), would produce a 3.6\,cm flux of
about 0.3\,mJy at Y\,Gem's distance.

\section{The Role of ngVLA}
The discovery of the large and variable UV and X-ray fluxes, and subsequent detection of radio emission from 
Y\,Gem supports our model with episodic accretion in a binary system for the high energy emission from fuvAGB stars. 
Although X-ray observations (with XMM-Newton and Chandra) and UV observations (with HST) are valuable probes of 
binary-associated accretion 
activity in AGB stars, our studies show that these are sensivity-limited to the brightest few ($\lesssim10$) objects in 
the full sample.

The ngVLA's sensitivity is needed to carry out a 
survey of a statistical sample of fuvAGB stars over the $\sim$3-90\,Ghz range (see below), and 
to probe their short and long-term variability. Such a survey will enable 
us to search for and investigate the nature of radio emission from fuvAGB stars. Accretion-related radio emission is 
expected to have both thermal and non-thermal components and display significantly time-variable on short time-scales 
(minutes to weeks), whereas chromospheric emission is expected to be thermal, possibly with time-variability, but only on 
long time-scales (many months to a year).

The long-term UV variability time-scale in fuvAGB stars ($\sim10^7$\,sec in Y\,Gem) is likely related to the semi-regular variability of the primary (the optical 
light curve of Y\,Gem shows 3 periodicities at 280\,d, 158\,d and 172\,d) due to its pulsations producing variations in the 
accretion mass-flux. We find variability on a medium time-scale (of a few days) in the H$\alpha$ profile of Y\,Gem 
suggesting variations in the inner regions (few$\times$100\,AU) of a fast outflow (e.g., a jet powered by the accretion 
disk).  The shortest variability timescale of $\sim20$\,s is due to the flickering phenomenon that characterizes active 
accretion disks. The wide frequency coverage provied by the ngVLA will enable detailed modeling the radio emission, and 
decomposition into thermal and non-thermal components. 

Using the ngVLA's predicted continuum performance as given by Selina et al. (2018), we find that with 15\,min  
integration time per source, we can achieve a 5$\sigma$ sensitivity of $\lesssim$2.5\,$\mu$Jy (4.8\,$\mu$Jy) for any ngVLA band 
in the 8-41\,GHz (90\,GHz) frequency range. Since this is a factor $> few \times 100$ lower than 
Y\,Gem's flux at $>10$\,GHz, we can carry out a large multi-band survey of several 100 AGB stars with UV emission with 
a modest expenditure of ngVLA time. Such a project cannot be done with any other facility: it is prohibitive in terms of time for 
the VLA even for the mere detection of AGB stars at the few\,$\mu$Jy level, and the VLA certainly lacks the sensitivity to probe  
the short-term (i.e., over minutes to hours) variability of these objects. 

The brightest (say) 20 of these would then define a key 
subsample for detailed follow-up time-monitoring studies. Since the ngVLA covers the low-J lines of several molecular 
species that are typically detected in the winds from AGB stars (e.g., CO, $^{13}$CO, HCN, SiO, CS), a pilot survey of 
these lines in the key subsample, together with ALMA observations of high-J lines, would help to probe the gas 
mass-loss rates and kinematics of the inner regions of outflows, and the presence of torii and/or disks in these stars 
resulting from the binary interaction (e.g., Sahai et al. 2017a,b). 

\acknowledgements RS gratefully acknowledges the contributions of his colleagues Carmen S{\'a}nchez Contreras, Jorge 
Sanz-Forcada, Mark Claussen, and C. Muthumariappan to the UV, X-ray, optical and radio studies described here, and 
discussions with Orsola de Marco, Eric Blackman, Albert Zijlstra, 
Joel Kastner, and Bruce Balick (in no particular order) during numerous proposal preparations, that have been 
helpful in motivating 
the study described here.


\end{document}